\documentclass[12pt]{article}
\usepackage[OT1]{fontenc} 
\usepackage[utf8]{inputenc}
\usepackage{amsmath}
\usepackage{graphicx}
\usepackage{enumerate}
\usepackage{natbib}
\usepackage{url}

\usepackage{amsthm}
\usepackage{amsfonts}
\usepackage{dsfont}
\usepackage{graphicx,graphics}
\usepackage{wrapfig}
\usepackage{enumitem}
\usepackage{booktabs}
\usepackage{threeparttable}
\usepackage{nicefrac} 
\usepackage{yfonts}
\usepackage[euler]{textgreek}
\usepackage[normalem]{ulem}
\usepackage{listings}
\usepackage{MnSymbol}
\usepackage{silence}
\usepackage{caption}
\usepackage{subcaption}
\usepackage{float}
\DeclareCaptionFormat{graybox}{{\setlength{\fboxsep}{0.1pt}\colorbox{gray!10}{\parbox{\dimexpr\linewidth-2\fboxsep}{#1#2#3}}}}
\DeclareCaptionFormat{grayboxindent}{{\setlength{\fboxsep}{0.1pt}\colorbox{gray!10}{\parbox{\dimexpr\linewidth-2\fboxsep}{\hspace{0.3em}#1#2#3}}}}

\newcommand{\panelbox}[1]{%
  \fboxsep=2pt\fboxrule=0.4pt%
  \fcolorbox{gray!50}{white}{#1}%
}

\usepackage{bm, verbatim}
\usepackage[english]{babel}
\usepackage{algorithmic}
\makeatletter
\def\thm@space@setup{%
  \thm@preskip=\parskip \thm@postskip=0pt
}
\makeatother
\usepackage{xr-hyper}
\usepackage{hyperref}
\IfFileExists{1-supplement.aux}{\externaldocument{1-supplement}}{}
\makeatletter
\newcommand{\sref}[1]{%
  \@ifundefined{r@#1}{\suppfb{#1}}{\ref*{#1}}%
}
\makeatother
\newcommand{\suppfb}[1]{%
  \ifcsname suppval@#1\endcsname\csname suppval@#1\endcsname\else ??\fi
}
\newcommand{\defsuppval}[2]{\expandafter\def\csname suppval@#1\endcsname{#2}}
\defsuppval{appendix:questionnaire}{A}
\defsuppval{appendix:learning_details}{B}
\defsuppval{app:pairwise_comparison}{B.4}
\defsuppval{appendix:utility_details}{C}
\defsuppval{app:f_j_construction}{C.2}
\defsuppval{app:cobb_douglas}{C.1}
\defsuppval{app:non_disclosure}{C.3}
\defsuppval{app:additional_sim_results}{C.4}
\defsuppval{fig:candidate_welfare}{C.1a}
\defsuppval{fig:candidate_welfare_unconditional}{C.1b}
\defsuppval{fig:interview_heatmap}{C.2}
\defsuppval{fig:blocking_pairs}{C.3}
\defsuppval{app:theoretical_details}{D}
\defsuppval{app:candidate_side}{D.1}
\defsuppval{app:monotonicity}{D.1.1}
\defsuppval{app:proofs_candidate}{D.1.2}
\defsuppval{app:approx_ic}{D.1.3}
\defsuppval{app:candidate_welfare}{D.1.4}
\defsuppval{app:department_side}{D.2}
\defsuppval{app:proofs_welfare}{D.2.1}
\defsuppval{app:proofs_stability}{D.2.2}
\defsuppval{app:offer_equilibrium}{D.2.3}
\defsuppval{app:proofs_two_rose}{D.2.4}
\defsuppval{lem:mono_offer}{D.2}
\defsuppval{thm:cand_welfare_gain_formal}{D.1}
\defsuppval{prop:finite_stability}{D.1}
\usepackage[linesnumbered, ruled, vlined]{algorithm2e}

\def\P{\mathbb{P}}
\def\E{\mathbb{E}}

\def\star{*}

\usepackage{xcolor}
\usepackage[textsize=tiny]{todonotes}

\newcounter{counter}[section]

\newtheorem{assumption}{Assumption}

\newtheorem{theorem}{Theorem}

\newcommand{\blind}{0}

\begin{document}

\def\spacingset#1{\renewcommand{\baselinestretch}%
{#1}\small\normalsize} \spacingset{1}


\if1\blind
{
\title{
\Large{\textbf{
{A Statistical Market-Design Framework for Academic Job Markets}
}}}

\author{
\bigskip
{\sc A and B} \\
{\it {\normalsize  University of C}}}
\date{}
\maketitle

} \fi

\if0\blind
{
  \title{\bf A Statistical Market-Design Framework for Academic Job Markets}
  \author{Ali Kaazempur-Mofrad\textsuperscript{1}, Xiaowu Dai\textsuperscript{1,2,$*$}, and Xuming He\textsuperscript{3}\\\\
  \textsuperscript{1} \it \normalsize Department of Statistics and Data Science, University of California, Los Angeles\\
  \textsuperscript{2} \it \normalsize Department of Biostatistics, University of California, Los Angeles\\
  \textsuperscript{3} \it \normalsize Department of Statistics and Data Science, Washington University in St. Louis}
  \date{}
  \maketitle
}\fi

\if0\blind
{
\begin{footnotetext}[1]
{\textit{Address for correspondence:} Xiaowu Dai, Department of Statistics and Data Science, UCLA, 8917 Math Sciences Bldg \#951554,  Los Angeles, CA 90095,  USA. Email: daix@ucla.edu.}
\end{footnotetext}
} \fi

\begin{abstract}
\noindent
The academic job market for new statisticians is highly congested at the interview stage, where departments must rank and select candidates from large applicant pools without credible signals of candidate interest. As a result, interviews and offers are often misallocated, leading to unfilled positions and poor mutual fit. We frame interview allocation as a statistical ranking problem under uncertainty and propose a market-design framework that incorporates structured preference signaling into interview selection.
Candidates submit a single standardized questionnaire describing preferences over interpretable job characteristics, which departments combine with traditional application materials and historical hiring data to estimate candidate-specific acceptance probabilities and expected utilities. To account for estimation uncertainty, we employ a confidence-calibrated ranking procedure based on pairwise utility comparisons that provides statistical guarantees for candidate ranking. We establish that truthful participation is optimal for candidates and that preference information improves departmental outcomes and matching stability. We use a dataset of U.S. statistics departments to show that the proposed framework substantially increases matching rates, improves match quality, and reduces hiring failures relative to the current practice. 
\end{abstract}

\bigskip
\noindent%
\vfill

\spacingset{1.8}

\section{Introduction}
\label{sec:intro}
The academic job market for new statisticians is characterized by severe congestion at the interview stage. Each year, hundreds of Ph.D. graduates and postdocs apply broadly across institutions for tenure-track positions, while departments must allocate a small number of interview slots. Although hiring decisions ultimately depend on mutual preferences, existing application materials, such as curricula vitae, research statements, and letters of recommendation, primarily convey information about candidate scholarly quality and provide little credible indication of candidate interest in specific positions. As a result, interview opportunities and offers are frequently misallocated: \emph{a small subset of candidates receive multiple interviews and offers but can accept only one position, while many departments fail to hire even after extensive searches}.

This imbalance is evident in recent data. In the 2023–24 academic year, 641 students earned Ph.D. degrees in statistics in the United States, yet only 82 accepted permanent academic positions, including tenure-track jobs, while 185 had not secured any post-graduation position \citep{SED2024}. At the same time, departments posted 130 assistant professor openings in statistics on MathJobs.org, 115 of which were tenure-track, but only 88 positions were filled, leaving 42 searches unfilled; the detailed list of these positions is provided in the supplementary material. Figure~\ref{fig:realmarket_inefficiency} illustrates the simultaneous failures on both sides of the market.
Because candidates can submit dozens of applications at low marginal cost, departments often receive far more applications than they can reasonably interview. As a result, competition for a limited number of interview slots becomes intense, generating substantial congestion and likely wasted effort on both sides of the market.

From a statistical perspective, interview selection in academic hiring is a ranking problem under uncertainty. Departments must rank candidates not only by scholarly quality but also by the probability that an offer will be accepted, a quantity that is rarely observed directly and must be inferred from application materials and the department’s historical hiring data. For many departments, ranking candidates solely by scholarly quality can be suboptimal, as highly accomplished candidates are often unattainable because they are likely to accept other jobs they prefer.
Reasonable and informative estimation of acceptance probabilities therefore plays a central role in effective interview allocation. However, without credible preference signals, departments face substantial uncertainty about candidate interest and may underestimate the acceptance probabilities of candidates who would be strong matches. 
This motivates the development of methods that enable credible signaling of candidate preferences and reduce coordination failures in which departments and candidates who would be well-suited fail to match. 

We propose a novel statistical market-design framework that introduces structured preference signaling into the academic job market. The framework augments existing application systems with a centralized mechanism in which each candidate may submit a single standardized questionnaire, for example, through platforms such as MathJobs, Interfolio, or the ASA website, to signal preferences over interpretable job characteristics such as geographic location, research–teaching balance, and departmental environment.
The \emph{single-questionnaire} constraint preserves the scarcity and credibility of the signal while enabling departments to interpret candidate preferences consistently. Departments combine questionnaire responses with traditional application materials and historical offer–acceptance data to estimate candidate-specific acceptance probabilities and expected utilities.
Building on these estimates, we develop a confidence-calibrated ranking procedure for interview selection that explicitly accounts for estimation uncertainty by calibrating pairwise differences in expected utility and providing statistically valid guarantees on candidate ranking. Overall, this framework provides a practical and statistically principled approach for allocating limited interview slots.

We analyze the properties of the proposed framework from both candidate and departmental perspectives. On the candidate side, we establish that participation and truthful reporting are optimal compared to alternative strategies such as non-participation and misreporting.  On the department side, we show that preference signaling improves welfare by enabling more effective allocation of interview slots. 
Moreover, as the questionnaire becomes more informative, the mechanism improves matching stability by reducing mismatches between candidates and departments. For example, without preference information, a department may extend an offer to a highly ranked candidate who ultimately prefers another institution, while overlooking a slightly lower-ranked candidate who would strongly prefer and accept the position. With informative preference signals, the department can more effectively identify candidates with genuine interest and prioritize those most likely to accept and remain.
This improved stability lowers the likelihood that candidates leave shortly after joining and reduces the need for costly repeated hiring searches.

To evaluate the practical impact of the proposed framework, we construct a new dataset characterizing U.S. statistics departments using publicly available U.S. News rankings and institutional data, and conduct numerical studies calibrated to realistic market conditions. Results over a ten-year horizon demonstrate substantial efficiency gains. Under full candidate participation, total candidate welfare increases by $69\%$, driven by higher matching rates, which rise from $8.0\%$ to $12.6\%$, corresponding to an increase from approximately $24$ to $38$ successful placements per annual cohort of $300$ candidates, as well as improved match quality. At the individual level, participation strictly dominates non-participation, with early adopters achieving over twelve times higher welfare than non-participants. On the department side, the gains are heterogeneous across prestige levels. Lower- and mid-tier institutions experience the largest improvements in fill rates, rising from $35.3\%$ to $58.2\%$ for mid-ranked departments and from $31.3\%$ to $54.9\%$ for the lowest-ranked departments, while upper-mid-ranked departments improve from $43.8\%$ to $67.6\%$ and the most prestigious departments from $70.8\%$ to $82.2\%$. These improvements demonstrate that the proposed framework substantially reduces hiring failures and improves match quality through improved identification of mutually preferred matches.

While motivated by the statistics job market, the proposed market-design framework is broadly applicable to other fields, including mathematics, computer science, and the social sciences, that rely on decentralized academic hiring processes in which departments make interview and hiring decisions independently and lack a shared mechanism for coordinating candidate preferences. By focusing on a set of interpretable job characteristics rather than detailed knowledge of individual departments, the framework reduces reliance on informal information and personal connections in expressing candidate preferences.  More generally, this framework demonstrates how market design can be integrated with statistical inference to address coordination failures in decision-making settings.

The remainder of the paper is organized as follows. Section~\ref{sec:marketdesign} introduces the market design framework and the dataset of U.S. statistics departments. Section~\ref{sec:selection} develops the learning framework and interview selection procedure. Section~\ref{sec:theoretical_guarantees} establishes theoretical properties of the mechanism, including candidate incentives for participation and truthful reporting, departmental welfare gains, matching stability, and comparison to alternative mechanisms. Section~\ref{sec:simulations} presents numerical studies and outlines a pilot implementation. Section~\ref{sec:relatedworks} reviews related work in job market design, preference signaling, and matching markets. Section~\ref{sec:conclusion} concludes the paper with future directions.

\begin{figure}[p]
\captionsetup[subfigure]{labelformat=simple,labelsep=space,font={bf,small},justification=raggedright,singlelinecheck=false,format=graybox}
\renewcommand{\thesubfigure}{(\alph{subfigure})}
\centering

\newlength{\tophalf}\setlength{\tophalf}{\dimexpr0.52\textheight-9.6em\relax}%
\newlength{\bothalf}\setlength{\bothalf}{\dimexpr0.48\textheight-9.6em\relax}%
\begin{subfigure}[t]{\textwidth}
    \caption{U.S.\ statistics job market (2023--24)}
    \label{fig:realmarket_inefficiency}
    \centering
    \panelbox{\includegraphics[width=\dimexpr\textwidth-8pt\relax,height=\dimexpr0.35\tophalf\relax]{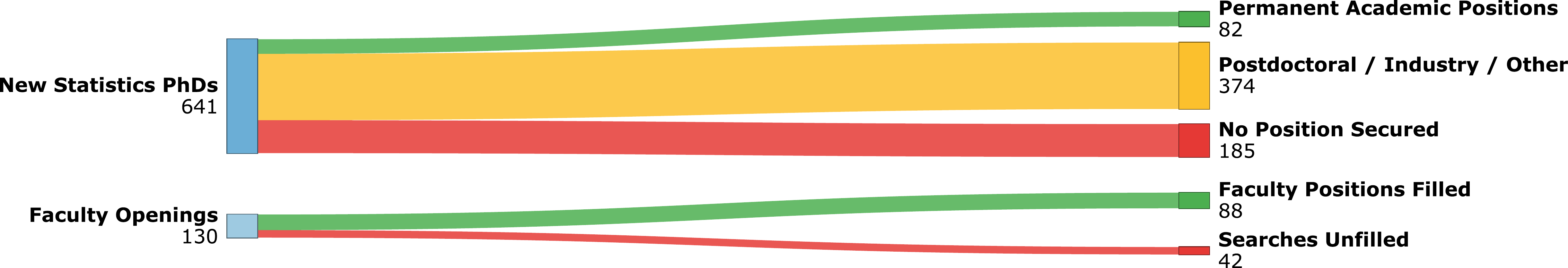}}
\end{subfigure}

\vspace{-0.3em}

\begin{minipage}[t]{0.46\textwidth}
    \setcounter{subfigure}{1}
    \begin{subfigure}[t]{\textwidth}
        \captionsetup{format=grayboxindent}
        \caption{Market-design framework}
        \label{fig:market_pipeline}
        \centering
        \panelbox{\includegraphics[width=\dimexpr\textwidth-5pt\relax,height=\dimexpr0.70\tophalf\relax]{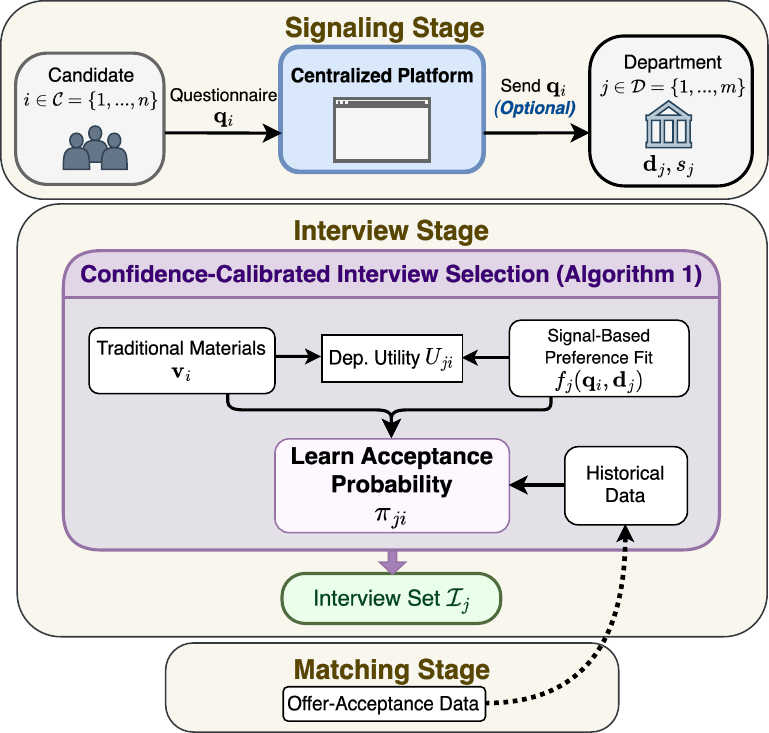}}
    \end{subfigure}
\end{minipage}%
\hfill
\begin{minipage}[t]{0.54\textwidth}
    \setcounter{subfigure}{2}
    \begin{subfigure}[t]{\textwidth}
        \caption{Interview selection}
        \label{fig:overview_ranking}
        \centering
        \panelbox{\includegraphics[width=\dimexpr\textwidth-5pt\relax,height=\dimexpr0.70\tophalf\relax]{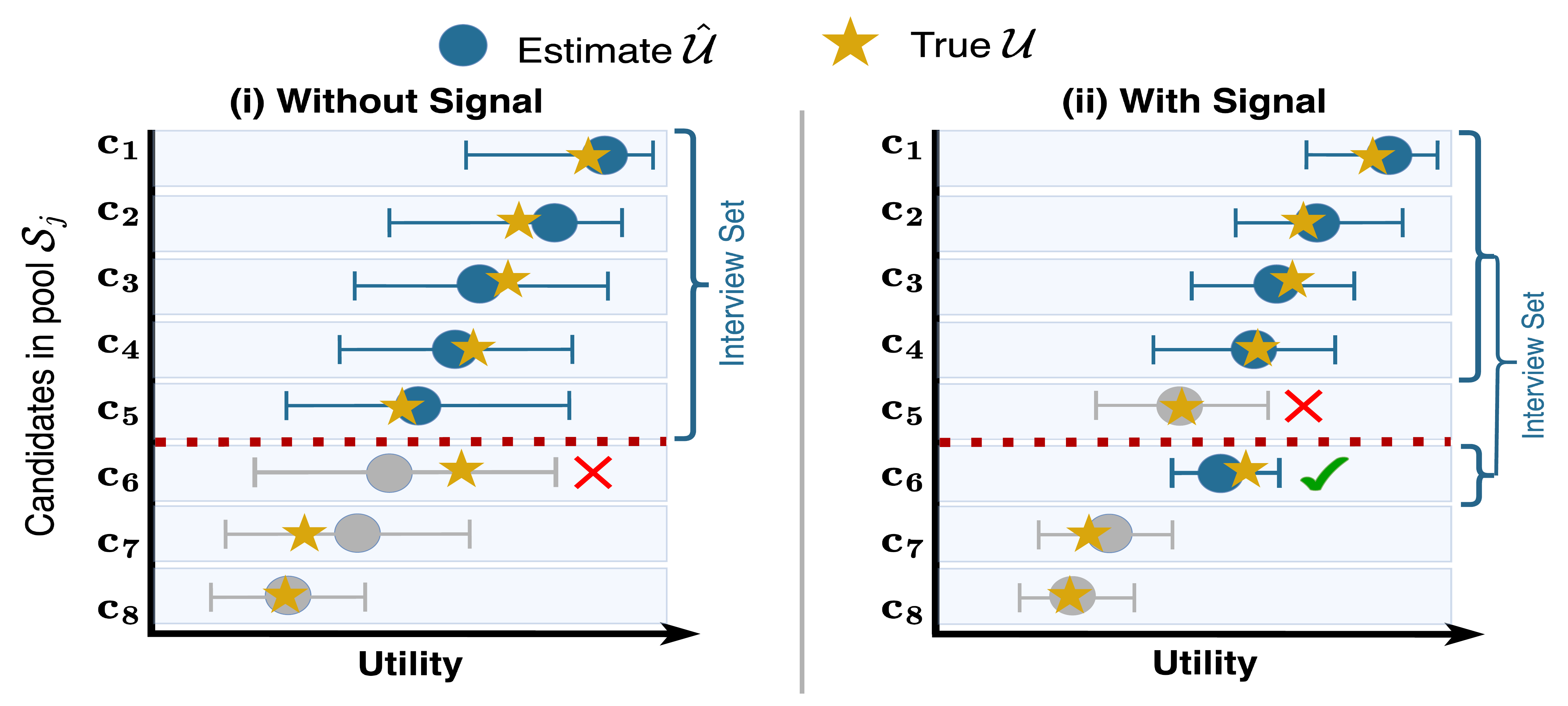}}
    \end{subfigure}
\end{minipage}

\vspace{-0.3em}

\setcounter{subfigure}{3}
\begin{subfigure}[t]{\textwidth}
    \caption{Proposed pilot implementation}
    \label{fig:overview_pilot}
    \centering
    \panelbox{\parbox[c]{\dimexpr\textwidth-8pt\relax}{\centering\includegraphics[width=0.9\textwidth,height=\bothalf]{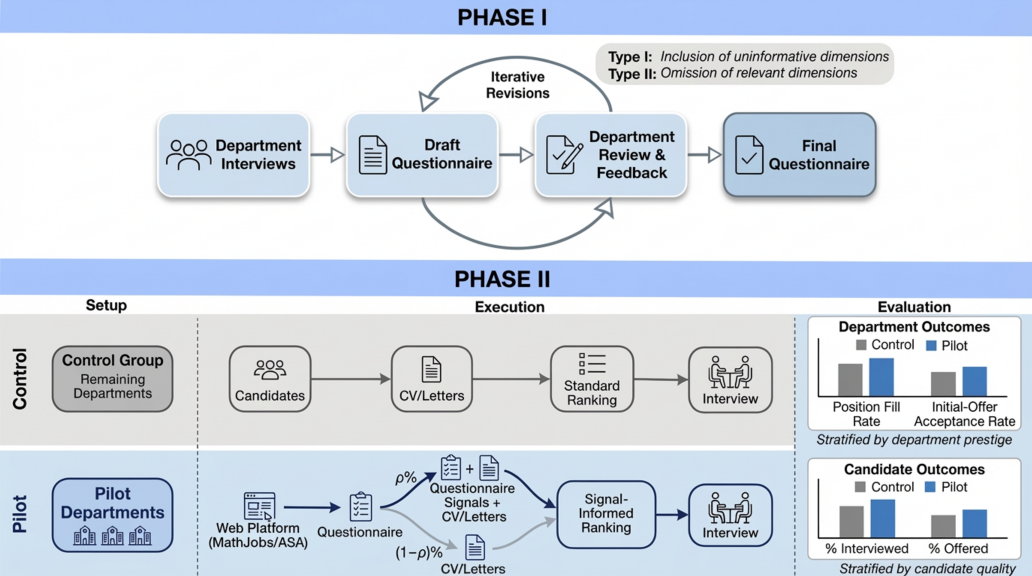}}}
\end{subfigure}

\caption{
(a)~Inefficiencies and missed allocations in the 2023--24 statistics job market (see Section~\ref{sec:intro}).
(b)~Three-stage market-design framework with preference signaling (see Section~\ref{sec:designjobmarket}).
(c)~Preference signals reduce estimation uncertainty and improve candidate ranking (see Section~\ref{subsec:confidence_calibrated_selection}).
(d)~Proposed pilot implementation (see Section~\ref{subsec:pilot}).}
\label{fig:overview}
\end{figure}
  
\section{Statistical Market-Design Framework}
\label{sec:marketdesign}

\subsection{Description of Dataset and Application}
\label{subsec:data_application}
We construct a dataset by integrating information from multiple available sources to characterize U.S. statistics departments. The primary source is the \citet{usnews2026statistics} graduate program rankings, which provide peer assessment scores and geographic information for 103 departments nationwide. We augment these data with institutional characteristics from the \citet{collegescorecard} College Scorecard. The resulting dataset provides structured and interpretable attributes of departments, including geographic location, perceived reputation, and institutional characteristics.

We use this dataset to construct a realistic setting for evaluating the proposed framework in the statistics job market. Each department is characterized by attributes drawn from the ranking and scorecard data. Candidates are then simulated with heterogeneous preferences over these attributes, as elicited through the standardized questionnaire described in Supplement~\sref{appendix:questionnaire}. This setup allows us to generate multiple years of historical offer–acceptance outcomes, which are used to estimate acceptance probabilities and inform interview selection.

Our application focuses on two central questions motivated by the academic job market for new statisticians:
\begin{itemize}[nosep]
\item[(Q1)] How can a simple and practical mechanism be designed to allocate interview slots more effectively?
\item[(Q2)] How can historical data support interview selection under uncertainty?
\end{itemize}

\subsection{Market-Design Framework with Signaling Mechanism}
\label{sec:designjobmarket}
We model the academic job market as a two-sided matching market with a set of hiring departments, denoted by $\mathcal{D} = \{1, \dots, m\}$, and a set of job candidates, denoted by $\mathcal{C} = \{1, \dots, n\}$. Each department $j \in \mathcal{D}$ has an interview capacity of $k_j$ candidates per hiring cycle. Departments are characterized by attributes $\mathbf{d}_j \in \mathbb{R}^{p_\mathrm{d}}$, such as geographic location, and by a parameter $s_j \in [0,1]$, where larger $s$ indicates higher reputation within the community. Because the marginal cost of submitting an additional application is low, we assume that candidates apply to all $m$ departments.

We propose a three-stage market-design framework for the academic job market, illustrated in Figure~\ref{fig:market_pipeline}:
\begin{itemize}
\item[(i)] \emph{Signaling stage.} Each candidate completes a single standardized questionnaire through a centralized platform, such as MathJobs, Interfolio, or the ASA website. The questionnaire elicits structured preferences over job characteristics, including geographic location. When submitting applications, candidates may choose whether to share the questionnaire with each department. If candidate $i$ elects to disclose the questionnaire response $\mathbf{q}_i\in \mathcal{Q}^{p_\mathrm{d}}$ to department $j$, the centralized platform transmits it directly to that department.
\item[(ii)] \emph{Interview stage.} Departments combine the questionnaire signals with traditional application materials, such as curricula vitae and letters of recommendation, to select candidates for interviews.
\item[(iii)] \emph{Matching stage.} The final stage follows standard hiring practice: departments extend offers to selected candidates, and candidates choose among available offers.
\end{itemize}
The questionnaire introduces a common signaling mechanism that is the novel component of the proposed market-design framework. The signal is both scarce and credible: each candidate completes the questionnaire once, but may decide whether to disclose it to any given department. We denote candidate $i$'s true preferences by $\mathbf{q}_i^* \in \mathcal{Q}^{p_\mathrm{d}}$ and the reported questionnaire response by $\mathbf{q}_i \in \mathcal{Q}^{p_\mathrm{d}}$, so that $\mathbf{q}_i = \mathbf{q}_i^*$ under truthful reporting. Throughout, departments observe and act on the reported preferences~$\mathbf{q}_i$.

\subsection{A Model of Department Utility}
\label{subsec:department_util}
Departments differ substantially in how they evaluate candidates. 
We model departmental evaluation using two sets of candidate features. 
The first is a quality vector $\mathbf{v}_i \in [0,1]^{p_\mathrm{v}}$, summarizing information from traditional application materials such as research output, teaching record, and letters of recommendation. 
The second is an alignment function $f_j(\mathbf{q}_i, \mathbf{d}_j) \in [\tfrac{1}{2},\,1]$, which measures how closely candidate $i$'s preferences, expressed through the questionnaire $\mathbf{q}_i$, align with the characteristics $\mathbf{d}_j$ of department $j$. The lower bound of $1/2$ reflects that even the least aligned candidate maintains a baseline level of compatibility with any department. This specific choice of $1/2$ is without loss of generality: since the framework depends only on relative ranking across candidates, any positive lower bound strictly less than $1$ would preserve the model properties and 
theoretical results.

Department $j$'s utility from hiring candidate $i$ is
\begin{equation}
\label{utility}
U_{ji} \;=\; U_j\big(\mathbf{v}_i,\, f_j(\mathbf{q}_i, \mathbf{d}_j)\big),
\end{equation}
where $U_j : [0,1]^{p_\mathrm{v}} \times [\tfrac{1}{2},\,1] \to [0,1]$ is a department-specific utility function. 
Examples of functional forms for $U_j$ used in the numerical studies are provided in Supplement~\sref{appendix:utility_details}. 
Regularity conditions on $U_j$ and $f_j$ are stated in Assumptions~\ref{assump:Uj} and~\ref{assump:fj}, which are standard in market and mechanism design \citep{azevedo2016}.

\begin{assumption}
\label{assump:Uj}
For each department $j$, the utility function $U_j$ is continuous and normalized such that
$U_j(\mathbf{0}, f) = 0$ for all $f \in [\tfrac{1}{2},1]$ and $U_j(\mathbf{1}, 1) = 1$.
It increases in $\mathbf{v}$ and strictly increases in
$f_j(\mathbf{q}_i, \mathbf{d}_j)$ whenever $\mathbf{v} \neq \mathbf{0}$.
\end{assumption}

\begin{assumption}
\label{assump:fj}
For each department $j$, the alignment $f_j(\mathbf{q}_i, \mathbf{d}_j)$ is non-degenerate; that is, there exist candidates $i$ and $k$ such that 
$f_j(\mathbf{q}_i, \mathbf{d}_j) \neq f_j(\mathbf{q}_k, \mathbf{d}_j)$.
\end{assumption}

Candidates may choose whether to disclose their questionnaire responses when submitting applications. If department $j$ receives the questionnaire from candidate $i$, it evaluates the candidate using both  $\mathbf{v}_i$ and $f_j(\mathbf{q}_i,\mathbf{d}_j)$ as specified in \eqref{utility}. 
If the questionnaire is not disclosed to department $j$, the department imputes the alignment score using a non-disclosure rule $\underline{f}_j$, defined as the minimum alignment score observed among candidates who disclosed their questionnaire to that department:
\begin{equation}
\label{eq:nondisclosure_rule}
\underline{f}_j \;=\; \min_{i' \in \mathcal{C}_j^{\mathrm{disc}}} 
f_j(\mathbf{q}_{i'}, \mathbf{d}_j),
\end{equation}
where $\mathcal{C}_j^{\mathrm{disc}} = \{i' \in \mathcal{C} : i' \text{ disclosed to department } j\}$. 
This pessimistic imputation rule follows the skeptical beliefs principle from the voluntary disclosure literature: non-disclosure is interpreted as the worst possible type consistent with silence \citep{milgrom1981good, hagenbach2014certifiable}. It ensures a fairness property: a candidate who withholds the questionnaire cannot receive a more favorable alignment evaluation than any candidate who discloses. The department therefore evaluates a non-disclosing candidate as
$U_{ji} = U_j(\mathbf{v}_i,\underline{f}_j)$.

\section{Confidence-Calibrated Ranking of Candidates}
\label{sec:selection}

\subsection{Ranking of Candidates}
\label{sec:rankingcandidates} 
The signaling mechanism in Section~\ref{sec:designjobmarket} provides additional information on candidate--department fit, which we incorporate into interview-stage decision-making through a ranking procedure. While the framework does not prescribe a unique ranking rule, we adopt a principled approach based on expected utility.
Let $\pi_{ji} \in [0,1]$ denote the candidate-specific probability that candidate $i\in\mathcal{C}$
accepts an offer from department $j\in\mathcal D$. The corresponding expected utility to
department $j$ from hiring $i$ is
\begin{equation}
\label{eqn:exputility}
\mathcal{U}_{ji} = U_{ji}\,\pi_{ji},
\end{equation}
where $U_{ji}$ is the departmental utility in \eqref{utility}.
Within department $j$, candidates are ranked by their expected
utilities in \eqref{eqn:exputility}. Specifically, the rank of candidate $i$ is defined as
\begin{equation*}
r_{ji} = \#\{\,\ell \neq i : \mathcal{U}_{j\ell} > \mathcal{U}_{ji} \,\} + 1,
\end{equation*}
so that candidates with larger expected utility receive better ranks.

\subsection{Learning Acceptance Probabilities}
\label{sec:learning}

We consider a learning framework to estimate the acceptance probabilities $\pi_{ji}$ in \eqref{eqn:exputility} using historical hiring data. Let $X_{ji,t}\in\{0,1\}$ indicate whether department $j$ extended an offer to candidate $i$ in year $t$, and let $Y_{ji,t}\in\{0,1\}$ indicate whether the offer was accepted.
The acceptance probability is defined as
\begin{equation*}
\pi_{ji}=\P\!\left(Y_{ji}=1 \,\middle|\, X_{ji}=1,\; s_j,\; \mathbf v_i,\;
f_j(\mathbf q_i,\mathbf d_j)\right),
\end{equation*}
that is, the probability that candidate $i$ accepts an offer from department $j$
conditional on the offer being made. 
We estimate $\pi_{ji}$ using a neural network with a penalized binary cross-entropy loss. Details of the network architecture and training procedures are provided in Supplement~\sref{appendix:learning_details}. 
Uncertainty in the estimated acceptance probabilities is quantified using the bootstrap \citep{efron1994}. Specifically, we refit the model on $B$ bootstrap resamples of the historical data, producing predictive draws $\{\pi_{ji}^{(b)}\}_{b=1}^B$. The bootstrap estimator is
\begin{equation}
\label{eqn:estpi}
\widehat{\pi}_{ji}=\frac{1}{B}\sum_{b=1}^B \pi_{ji}^{(b)},
\end{equation}
with corresponding expected-utility estimates
\begin{equation}
\label{eqn:hatu}
\mathcal{U}_{ji}^{(b)} = U_{ji}\pi_{ji}^{(b)},
\qquad
\widehat{\mathcal{U}}_{ji} = U_{ji}\widehat{\pi}_{ji}.
\end{equation}

\subsection{Confidence-Calibrated Ranking}
\label{subsec:confidence_calibrated_selection}
Based on the estimated acceptance probabilities $\pi_{ji}$ in \eqref{eqn:estpi}, we introduce a confidence-calibrated ranking rule that maximizes expected hiring outcomes while accounting for statistical uncertainty in candidate rankings.

\paragraph{Tier-based screening.}
Prestigious departments typically focus on a narrow pool of top candidates, whereas lower-ranked departments recruit more broadly \citep{clauset2015systematic, wapman2022quantifying}. To capture this pattern, we partition departments into $T$ tiers by prestige, $\mathcal{D} = \bigcup_{t=1}^{T} \mathcal{D}^{(t)}$, with lower indices corresponding to higher prestige. For each department $j \in \mathcal{D}^{(t)}$, we assume that candidate screening is restricted to a tier-based pool $\mathcal{S}_j = \bigcup_{r=1}^{t} \mathcal{C}^{(r)}$, so that higher-tier departments consider only top-tier candidates while lower-tier departments evaluate a broader set. The candidate tiers $\{\mathcal{C}^{(t)}\}_{t=1}^T$ are constructed recursively to match these screening pools: $\mathcal{C}^{(1)}$ consists of candidates considered by top-tier departments, and for $t \ge 2$, $\mathcal{C}^{(t)} = \mathcal{S}_j \setminus \bigcup_{r=1}^{t-1} \mathcal{C}^{(r)}$ for $j \in \mathcal{D}^{(t)}$. This construction ensures that $\mathcal{S}_j = \bigcup_{r=1}^{t} \mathcal{C}^{(r)} = \mathcal{C}^{(\le t)}$ for all $j \in \mathcal{D}^{(t)}$, and $\mathcal{C} = \bigcup_{t=1}^{T} \mathcal{C}^{(t)}$. 
All subsequent ranking and interview selection are conducted within $\mathcal{S}_j$.

\paragraph{Pairwise uncertainty in expected utilities.}
The goal is to construct an interview set that, with probability at least $1-\alpha$, contains every candidate whose true expected utility $\mathcal U_{ji}$ ranks among the top $k_j$. To quantify ranking uncertainty, we analyze all pairwise differences in expected utility. For each $(i,\ell)\in\mathcal{S}_j\times\mathcal{S}_j$, define
\begin{equation*}
\Delta_{j,i\ell} = \mathcal U_{ji} - \mathcal U_{j\ell}, \qquad
\widehat{\Delta}_{j,i\ell} = \widehat{\mathcal U}_{ji} - \widehat{\mathcal U}_{j\ell}.
\end{equation*}
Let $\Delta_{j,i\ell}^{(b)} = \mathcal U_{ji}^{(b)} - \mathcal U_{j\ell}^{(b)}$  denote the bootstrap draws defined in \eqref{eqn:hatu}, where $b=1,\ldots,B$. The estimated variance of the pairwise difference is
\begin{equation*}
\widehat{\sigma}^2_{j,i\ell}
=\frac{1}{B-1}\sum_{b=1}^B
\bigl(\Delta_{j,i\ell}^{(b)} - \widehat{\Delta}_{j,i\ell}\bigr)^2.
\end{equation*}

\paragraph{Simultaneous calibration.}
To control ranking errors uniformly when comparing any two candidates $i,\ell$ within department $j$'s candidate pool, we construct the max statistic
\begin{equation*}
\label{eq:max_stat}
Z_j^{(b)}
=
\max_{i\neq \ell}
\frac{\Delta_{j,i\ell}^{(b)} - \widehat{\Delta}_{j,i\ell}}
{\widehat{\sigma}_{j,i\ell}},
\end{equation*}
and let $c_{1-\alpha}^{\circ}$ denote the $(1-\alpha)$th quantile of $\{Z_j^{(b)}\}_{b=1}^B$. 
This yields the confidence set
\begin{equation}
\label{eq:CI_pairwise}
\mathcal R_{1-\alpha}=\Bigl\{\Delta:\;|\Delta_{j,i\ell}-\widehat{\Delta}_{j,i\ell}|\le c_{1-\alpha}^{\circ}\widehat{\sigma}_{j,i\ell},\ \forall i\neq \ell\Bigr\}.
\end{equation}
If the bootstrap procedure in Section~\ref{sec:learning} consistently approximates the joint distribution of $\widehat{\Delta}_{j,i\ell}$, then the confidence set in \eqref{eq:CI_pairwise} satisfies the coverage $\mathbb P\!\left(\{\Delta_{j,i\ell}\}_{i\neq \ell}\in\mathcal R_{1-\alpha}\right)\ge 1-\alpha$ \citep{mogstad2024inference}.

Figure~\ref{fig:overview_ranking} illustrates how preference signals affect the confidence set. Without the questionnaire, acceptance probabilities $\widehat{\pi}_{ji}$ are estimated using only candidate quality $\mathbf{v}_i$, leading to substantial uncertainty among similarly qualified candidates. This results in wide confidence intervals for $\widehat{\Delta}_{j,i\ell}$, making it difficult to distinguish candidates near the interview cutoff $k_j$. 
Incorporating the alignment score $f_j(\mathbf{q}_i, \mathbf{d}_j)$ allows departments to differentiate candidates based on fit, which reduces uncertainty in $\widehat{\pi}_{ji}$ and hence in $\widehat{\sigma}_{j,i\ell}$. 

\begin{algorithm}[t]
\caption{Confidence-Calibrated Interview Selection for Department $j$}
\label{alg:confidence_calibrated_selection}
\begin{algorithmic}[1]
\STATE \textbf{Input:} Historical offer–acceptance data, the current candidate list $\mathcal{S}_j$ and each candidate’s application materials, and the interview budget $k_j \in \mathbb{N}$.
\STATE \textbf{Step 1:} Compute the acceptance probability $\widehat{\pi}_{ji}$ by \eqref{eqn:estpi} and the expected utility $\widehat{\mathcal U}_{ji} = U_{ji}\widehat{\pi}_{ji}$ for all $i \in \mathcal{S}_j$ by \eqref{eqn:hatu}.
\STATE \textbf{Step 2:}  Compute the lower bound on each candidate $i$’s rank $\underline{R}_{ji}$ via \eqref{eq:rank_lower}, and let $\widehat{\mathcal I}_{k_j} = \{i \in \mathcal{S}_j : \underline{R}_{ji} \le k_j\}$. 
\STATE \textbf{Step 3:} \textbf{If} $|\widehat{\mathcal I}_{k_j}| \ge k_j$, select the top $k_j$ by $\widehat{\mathcal U}_{ji}$ within $\widehat{\mathcal I}_{k_j}$. 
Let $\mathcal I_j \leftarrow \widehat{\mathcal I}_{k_j}$.
\STATE \textbf{Output:} Interview set $\mathcal I_j$.
\end{algorithmic}
\end{algorithm}

\paragraph{Confidence-calibrated ranking and interview selection.}
The true rank of candidate $i$ within department $j$'s pool is
$r_{ji} = 1 + \sum_{\ell\neq i} \mathbf 1\{\Delta_{j,\ell i}>0\}$.
Since $\widehat{\Delta}_{j,\ell i}>c_{1-\alpha}^{\circ}\,\widehat{\sigma}_{j,\ell i}$ implies $\Delta_{j,\ell i}>0$ within $\mathcal R_{1-\alpha}$ by~\eqref{eq:CI_pairwise}, we obtain the lower bound,
\begin{equation}
\label{eq:rank_lower}
\underline{R}_{ji} = 1 + \sum_{\ell\neq i} \mathbf{1}\!\left\{\widehat{\Delta}_{j,\ell i} > c_{1-\alpha}^{\circ}\,\widehat{\sigma}_{j,\ell i} \right\}
\end{equation}
on $r_{ji}$. As $\underline{R}_{ji}\le r_{ji}$ holds simultaneously for all $i\in\mathcal{S}_j$, the interval $\left[\underline{R}_{ji},\ |\mathcal{S}_j|\right]$ is a simultaneous $(1-\alpha)$ one-sided confidence interval for $r_{ji}$, ensuring that
\begin{equation}
\label{eq:rank_guarantee}
\mathbb{P}\left(r_{ji} \geq \underline{R}_{ji} \text{ for all } i \in \mathcal{S}_j\right) \geq 1-\alpha.
\end{equation}
We then define the confidence-calibrated inclusion set
$\widehat{\mathcal I}_{k_j} =\{i\in\mathcal{S}_j : \underline{R}_{ji}\le k_j\},$
which, by \eqref{eq:rank_guarantee}, contains every truly top-$k_j$ candidate with probability at least $1-\alpha$.
If $|\widehat{\mathcal I}_{k_j}|\ge k_j$, the department selects the $k_j$ candidates within this set with the largest estimated expected utilities $\widehat{\mathcal U}_{ji}$. The resulting interview set is denoted by $\mathcal I_j$, with $|\mathcal I_j| = k_j$.

We summarize the confidence-calibrated ranking and interview selection procedure in Algorithm~\ref{alg:confidence_calibrated_selection}, which offers two advantages. First, unlike direct sorting by the plug-in estimates $\widehat{\mathcal U}_{ji}$, it explicitly accounts for heterogeneous uncertainty by calibrating each pairwise comparison $(\ell, i)$ using the estimated standard error $\widehat{\sigma}_{j,\ell i}$ in \eqref{eq:CI_pairwise}. Second, the procedure provides joint coverage guarantees for ranking for any candidate $i$ in \eqref{eq:rank_guarantee}. A detailed comparison with alternative ranking methods is given in Supplement~\sref{appendix:learning_details}.

\section{Theoretical Guarantees}
\label{sec:theoretical_guarantees}

\subsection{Incentives for Candidate's Participation}
\label{subsec:candidate_incentives}
A candidate considering the proposed mechanism faces two strategic choices: first, whether to disclose the questionnaire to departments, and second, what preferences to report. We show that disclosure is a dominant strategy: each candidate maximizes expected utility by participating and sharing the questionnaire with all departments, regardless of other candidates' actions. In addition, the benefits of misreporting preferences diminish under market competition, making truthful reporting optimal.

\paragraph{Candidate's Participation and Disclosure.}
Let $V_{ij} \in [0,1]$ denote candidate $i$'s utility from joining department $j$, so that $V_{ij} \ge V_{ik}$ whenever candidate $i$ prefers $j$ to $k$. Each candidate $i$ chooses a disclosure set $\mathcal{D}_i^{\mathrm{disc}} \subseteq \mathcal{D}$, representing the departments $\mathcal{D}_i^{\mathrm{disc}}$ to which the completed questionnaire is shared. 

Given questionnaire report $\mathbf{q}_i$ and disclosure set $\mathcal{D}_i^{\mathrm{disc}}$, candidate $i$'s expected payoff is
\begin{equation}
\label{eqn:defwic}
W_i^{\mathcal{C}}(\mathbf{q}_i,\, \mathcal{D}_i^{\mathrm{disc}})
=\sum_{j \in \mathcal{D}}V_{ij}\,P^{\mathrm{offer}}_{ji}
(\mathbf{q}_i, \mathcal{D}_i^{\mathrm{disc}}, \mathbf{v}_i;\,
\mathbf{q}_{-i}, \mathcal{D}_{-i}^{\mathrm{disc}}, \mathbf{v}_{-i}),
\end{equation}
where $\mathbf{q}_{-i}$, $\mathcal{D}_{-i}^{\mathrm{disc}}$, and $\mathbf{v}_{-i}$ denote the reports, disclosure decisions, and qualities of other candidates. The offer probability $P^{\mathrm{offer}}_{ji}$ depends on the department's estimated expected utility $\widehat{\mathcal U}_{ji}$ and therefore on candidate $i$'s strategy $(\mathbf q_i,\mathcal D_i^{\mathrm{disc}})$ through the alignment score $f_j(\mathbf q_i,\mathbf d_j)$.  The market participation rate $\rho$, defined as the fraction of candidates who submit the questionnaire, affects $P^{\mathrm{offer}}_{ji}$ through $\mathcal{D}_{-i}^{\mathrm{disc}}$, since non-participating candidates have $\mathcal{D}_{i'}^{\mathrm{disc}}=\emptyset$.
If candidate $i$ does not disclose to department $j$, the alignment score is imputed as $f_j(\mathbf{q}_i, \mathbf{d}_j)=\underline f_j$ according to the worst-discloser rule in  \eqref{eq:nondisclosure_rule}.

\begin{assumption}
\label{ass:uncertainty}
At the time of disclosure, candidate $i$ does not observe $f_j(\mathbf q_i,\mathbf d_j)$ for any $j\in\mathcal D$. Conditional on $\mathbf q_i$, the alignment scores $\{f_j(\mathbf q_{i'},\mathbf d_j)\}_{i'\in\mathcal C}$ are i.i.d.\ across candidates, and $\{f_j(\mathbf q_{i'},\mathbf d_j)\}_{j\in\mathcal D}$ are independent across departments.
\end{assumption}

\noindent
Assumption \ref{ass:uncertainty} reflects the information structure of the mechanism. 
The i.i.d.\ assumption across candidates captures that, conditional on $\mathbf{q}_i$, candidates have the same prior uncertainty and complete the questionnaire independently, yielding alignment scores that behave as independent draws from a common distribution. Independence across departments follows because evaluation criteria differ across institutions, so alignment with one department provides no information about alignment with another.

\begin{theorem}
\label{thm:disclosure}
Under Assumption~\ref{ass:uncertainty} and the rule~\eqref{eq:nondisclosure_rule}, universal disclosure is a dominant strategy in expectation: For any candidate $i$ and any disclosure set $\tilde{\mathcal D}_i^{\mathrm{disc}}\subseteq\mathcal D$, including selective disclosure $\tilde{\mathcal D}_i^{\mathrm{disc}}\subsetneq\mathcal D$ or non-participation $\tilde{\mathcal D}_i^{\mathrm{disc}}=\emptyset$, we have
$\E\!\left[W_i^{\mathcal C}(\mathbf q_i,\mathcal D)\right]
\ge\E\!\left[W_i^{\mathcal C}(\mathbf q_i,\tilde{\mathcal D}_i^{\mathrm{disc}})\right]$, where $W_i^{\mathcal C}$ is defined in~\eqref{eqn:defwic}.
\end{theorem}

\noindent
We make two remarks on Theorem~\ref{thm:disclosure}. First, the proof relies on first-order stochastic dominance \citep{shaked2007stochastic}: a candidate’s alignment score $f_j(\mathbf{q}_i, \mathbf{d}_j)$ stochastically dominates the worst-discloser imputation $\underline f_j=\min_{i'\in\mathcal C_j^{\mathrm{disc}}} f_j(\mathbf{q}_{i'}, \mathbf{d}_j)$. Since offer probabilities are nondecreasing in alignment, disclosure increases the expected offer probability and hence the candidate’s expected payoff.
Second, Theorem~\ref{thm:disclosure} implies that each candidate’s expected welfare is higher under disclosure, regardless of other candidates’ strategies. At the aggregate level, the mechanism reallocates offer probabilities toward candidate--department pairs with higher mutual fit, thereby increasing total candidate welfare. 

\paragraph{Candidate's Truthful Reporting.}
Theorem~\ref{thm:disclosure} shows that candidates should participate and disclose their questionnaire, but it does not rule out strategic misreporting. In competitive markets, however, the gains from such behavior are limited: highly desirable departments attract many strong candidates, so misreporting yields only small increases in offer probability, while the loss of alignment at less competitive departments can be substantial. We formalize this competitive market condition below.
\begin{assumption}\label{ass:competitive}
For any department tier $t$ and interview capacity $K = \sum_{j \in \mathcal{D}^{(t)}} k_j$:  
\begin{enumerate}[label=(\roman*),nosep]
\item \emph{Large pool.} Let $\mathcal{H} \subseteq \mathcal{C}^{(\le t)}$ satisfy $\|\mathbf{v}_i - \mathbf{v}_\ell\| \le \delta$ for all $i, \ell \in \mathcal{H}$, where $\mathcal{C}^{(\le t)} := \bigcup_{r=1}^{t} \mathcal{C}^{(r)}$. Define $n_c := |\mathcal{H}|$. We assume that $n_c > K$ to ensure that there are more candidates with comparable qualities than available interview slots.
\item \emph{Bounded dispersion.} Candidate utilities within $\mathcal{H}$ are uniformly close: $|U_j(\mathbf{v}_i,f) - U_j(\mathbf{v}_\ell,f)| \le L_U\delta$ for all $i,\ell \in \mathcal{H}$ and some $L_U>0$. 
\end{enumerate}
\end{assumption}

\noindent
Assumption \ref{ass:competitive}(i) requires that the number of candidates with comparable qualities exceeds the total interview capacity within each tier.
Assumption \ref{ass:competitive}(ii) follows if $U_j$ is $L_U$-Lipschitz. Since $\|\mathbf{v}_i - \mathbf{v}_\ell\| \le \delta$ for all $i,\ell \in \mathcal{H}$, this implies
$|U_j(\mathbf{v}_i,f) - U_j(\mathbf{v}_\ell,f)| \le L_U |\mathbf{v}_i - \mathbf{v}_\ell| \le L_U\delta$.
Thus, candidates in $\mathcal{H}$ have similar utilities, so alignment differences play a primary role in distinguishing among them.

\begin{theorem}
\label{thm:approx_ic_main}
Under Assumptions~\ref{ass:uncertainty}--\ref{ass:competitive},
for any candidate $i \in \mathcal{C}$ with true preferences $\mathbf{q}_i^*$, the gain from any misreport $\mathbf{q}_i'$ is bounded by,
$$W_i^{\mathcal{C}}(\mathbf{q}_i',\, \mathcal{D}) - W_i^{\mathcal{C}}(\mathbf{q}_i^*,\, \mathcal{D}) \;\le\; \varepsilon = L_U\delta \;+\; \frac{K}{n_c} \;+\; o(n_c^{-1}),$$
where $\varepsilon \to 0$ as $n_c / K \to \infty$ and $\delta \to 0$.
\end{theorem}
\noindent 
We make two remarks on Theorem \ref{thm:approx_ic_main}. First, it shows that as the ratio $n_c/K$ increases and $\delta$ decreases, the bound $\varepsilon$ converges to zero, so the gains from misreporting vanish in competitive markets. Together with Theorem~\ref{thm:disclosure}, this implies that disclosure and truthful reporting are optimal candidate strategies.
Second, both Theorems~\ref{thm:disclosure} and~\ref{thm:approx_ic_main} hold regardless of the participation rate $\rho$: disclosure and truthful reporting are independent of other candidates’ behavior, and the bound on misreporting gains depends only on market competitiveness rather than the overall participation rate, thereby encouraging each candidate to participate in the mechanism at any level of participation rate.

\subsection{Incentives for Department's Participation}
\label{subsec:department_incentives}
\noindent
A department considering the proposed mechanism faces two questions: first, whether the questionnaire improves interview selection, and second, whether the mechanism improves the stability of market outcomes. We show that preference information improves both expected departmental outcomes and stability as the preference signal becomes more informative.

\paragraph{Department's Gain from Preference Information.}
Without the questionnaire, departments select candidates based only on quality signals $\mathbf{v}_i$. The questionnaire augments these signals with alignment scores $f_j(\mathbf{q}_i, \mathbf{d}_j)$, allowing departments to distinguish among similarly qualified candidates based on mutual fit.

Since $\mathbf{q}_i \in \mathcal{Q}^{p_\mathrm{d}}$ and $\mathbf{d}_j \in \mathbb{R}^{p_\mathrm{d}}$ share a common dimension $p_\mathrm{d}$, we measure informativeness by the questionnaire dimension $p_\mathrm{d}$. A higher-dimensional questionnaire captures richer preference information and yields more informative alignment signals. Formally, we assume that a $p_\mathrm{d}'$-dimensional questionnaire with $p_\mathrm{d}' > p_\mathrm{d}$ is more informative, in the sense that the induced alignment scores $\{f_j(\mathbf{q}_i^{(p_\mathrm{d}')}, \mathbf{d}_j^{(p_\mathrm{d}')})\}_{i \in \mathcal{S}_j}$ generate a finer $\sigma$-algebra than $\{f_j(\mathbf{q}_i^{(p_\mathrm{d})}, \mathbf{d}_j^{(p_\mathrm{d})})\}_{i \in \mathcal{S}_j}$.
Department $j$'s expected payoff under questionnaire dimension $p_\mathrm{d}$ is
\begin{equation}
\label{eqn:defwjd}
W_j^{\mathcal{D}}(p_\mathrm{d}) \;=\; \E\left[\max_{\substack{\mathcal{I}_j \subseteq \mathcal{S}_j,\; |\mathcal{I}_j| = k_j}}\ \sum_{i \in \mathcal{I}_j} \E\left[\mathcal{U}_{ji} \;\middle|\; \{f_j(\mathbf{q}_i^{(p_\mathrm{d})}, \mathbf{d}_j^{(p_\mathrm{d})})\}_{i \in \mathcal{S}_j}\right]\right],\end{equation}
where $\mathcal{U}_{ji}$ is the expected utility in~\eqref{eqn:exputility}.

\begin{theorem}\label{thm:dept_marginal}
If the $p_\mathrm{d}'$-dimensional questionnaire is more informative than the $p_\mathrm{d}$-dimensional questionnaire ($p_\mathrm{d}' > p_\mathrm{d}$), in the sense that
$\sigma\big(\{f_j(\mathbf{q}_i^{(p_\mathrm{d})}, \mathbf{d}_j^{(p_\mathrm{d})})\}_{i \in \mathcal{S}_j}\big)\;\subseteq\; \sigma\big(\{f_j(\mathbf{q}_i^{(p_\mathrm{d}')}, \mathbf{d}_j^{(p_\mathrm{d}')})\}_{i \in \mathcal{S}_j}\big)$, then $W_j^{\mathcal{D}}(p_\mathrm{d}') \geq W_j^{\mathcal{D}}(p_\mathrm{d})$.
\end{theorem}
\noindent
We make two remarks on Theorem \ref{thm:dept_marginal}. First, by the classical value-of-information principle \citep{blackwell1953equivalent}, a more informative signal allows the decision rule to condition on finer information and thus improve the optimal payoff. 
Second, this result addresses whether the questionnaire improves interview selection: with more informative alignment signals, departments can better distinguish among candidates with similar quality and select those with higher expected utility, leading to a more effective allocation of interview slots. In practice, a richer questionnaire improves the accuracy of candidate ranking by reducing the variance of alignment estimates and increases acceptance rates by selecting better-matched candidates.

\paragraph{Stability of Mechanism Outcomes.}
A pair $(i,j)$ is a \emph{blocking pair} of a matching $\mu$ if department $j$ strictly prefers candidate $i$ to its matched candidate under $\mu$, and candidate $i$ strictly prefers $j$ to their matched department. Reducing blocking pairs is central to stability \citep{gale1962college}, as their presence creates incentives for candidates to leave shortly after joining, leading to renegotiation, turnover, and costly repeated hiring searches.

\begin{assumption}\label{ass:CI-limit}
Under truthful reporting, as the questionnaire dimension $p_\mathrm{d} \to \infty$, the alignment signal is consistent for the true alignment, i.e.,
$f_j(\mathbf{q}_i^{(p_\mathrm{d})}, \mathbf{d}_j^{(p_\mathrm{d})}) \xrightarrow{p} f_{ji}^\star\ := f_j(\mathbf{q}_i^*, \mathbf{d}_j^*)$, for all $i,j$,
where $\mathbf{q}_i^{(p_\mathrm{d})}$ and $\mathbf{d}_j^{(p_\mathrm{d})}$ denote the $p_\mathrm{d}$-dimensional questionnaire responses and department attributes, and $\mathbf{q}_i^*$ and $\mathbf{d}_j^*$ denote the corresponding true candidate preferences and department characteristics.
\end{assumption}

Since both the departmental utility $U_{ji}$ in~\eqref{utility} and the estimated acceptance probability $\widehat{\pi}_{ji}$ in~\eqref{eqn:estpi} depend on $f_j(\mathbf{q}_i, \mathbf{d}_j)$, any deviation from $f_{ji}^\star$ propagates into the estimated expected utilities $\widehat{\mathcal{U}}_{ji}$ in~\eqref{eqn:hatu}. Let $\mu_{p_\mathrm{d}}$ denote the matching produced under the questionnaire with dimension~$p_\mathrm{d}$, let $\mu^\star$ denote the matching under $f_{ji}^\star$ utilities $U_{ji}^\star = U_j(\mathbf{v}_i, f_{ji}^\star)$, and let $B(\mu)$ denote the number of blocking pairs of~$\mu$.

\begin{theorem}\label{thm:blocking_pairs}
Under Assumptions~\ref{assump:Uj} and~\ref{ass:CI-limit}, if $U_j$ is $L_U$-Lipschitz, then
\begin{equation*}
\E[B(\mu_{p_\mathrm{d}})]\;\le\;B(\mu^\star)\;+\;nm \sum_{j=1}^{m}\sum_{i \in \mathcal{S}_j}\P\!\left(\left|f_j(\mathbf{q}_i^{(p_\mathrm{d})}, \mathbf{d}_j^{(p_\mathrm{d})}) - f_{ji}^\star\right|\ge\frac{U_{\min}}{2L_U}\right),
\end{equation*}
where $U_{\min} := \min_{j'} \min_{i' \neq \ell \in \mathcal{S}_{j'}} |U_{j'i'}^\star - U_{j'\ell}^\star|$ and $U_{ji}^\star := U_j(\mathbf{v}_i, f_{ji}^\star)$. In particular, $\E[B(\mu_{p_\mathrm{d}})] \to B(\mu^\star)$ as $p_\mathrm{d} \to \infty$.
\end{theorem}

\noindent
We make two remarks on Theorem \ref{thm:blocking_pairs}. First, the term $B(\mu^\star)$ captures inherent instability under complete information $p_\mathrm{d}\to\infty$, while the term $\P(|f_j(\mathbf{q}_i^{(p_\mathrm{d})}, \mathbf{d}_j^{(p_\mathrm{d})}) - f_{ji}^\star|\ge U_{\min}/2L_U)$ quantifies additional instability due to incomplete information on alignment. Misranking occurs only when the difference $|f_j(\mathbf{q}_i^{(p_\mathrm{d})}, \mathbf{d}_j^{(p_\mathrm{d})}) - f_{ji}^\star|$ is large enough to overturn utility differences $\widehat{\Delta}_{j,i\ell} = \widehat{\mathcal U}_{ji} - \widehat{\mathcal U}_{j\ell}$ in \eqref{eq:CI_pairwise}; the threshold $U_{\min}/(2L_U)$ reflects the smallest utility margin across all candidate comparisons. As the questionnaire becomes more informative, this error probability  $\P(|f_j(\mathbf{q}_i^{(p_\mathrm{d})}, \mathbf{d}_j^{(p_\mathrm{d})}) - f_{ji}^\star|\ge U_{\min}/2L_U)$ vanishes.

\subsection{Comparison to Alternative Mechanisms}
\label{subsec:alt_mechanisms}
Several market-design approaches have been proposed to mitigate coordination failures in job markets. We compare the proposed framework with the most closely related mechanisms and highlight their key differences, as summarized in Table~\ref{tab:mechanism_comparison}.

\begin{table}[ht]
\centering
\begin{threeparttable}
\caption{Comparison of matching market mechanisms.}
\label{tab:mechanism_comparison}
\small
\begin{tabular}{@{}lccccc@{}}
\toprule
Feature & AEA Binary Signaling & Deferred Acceptance  & \textbf{Proposed} \\
\midrule
Signal type & Binary & Ranked list   & Multi-attribute \\
Dept.\ coverage & $\le 2$ & All   & All \\
Incentive compatible & Partial & Yes  & Yes \\
Decentralized market & Yes & No  & Yes \\
\bottomrule
\end{tabular}
\end{threeparttable}
\end{table}

The first is the American Economic Association (AEA) preference signaling mechanism, which allows each candidate to send a small number of binary signals (typically two) to indicate strong interest in selected departments \citep{coles2010job, coles2013preference}. While these signals are credible due to their scarcity, they convey limited information and are sent to only a small subset of departments.  As a result, the mechanism is only partially incentive compatible: candidates may strategically allocate signals (e.g., one to a top choice and one to a safer option) rather than to their most preferred departments.  Although it improves coordination relative to no signaling, it does not fully resolve market inefficiencies. In Supplement~\sref{app:theoretical_details}, we further show that the proposed questionnaire mechanism improves expected departmental welfare relative to the AEA signaling mechanism.

The second is the centralized mechanism that collects full ranked preference lists and computes stable matchings via the Nobel Prize–winning deferred acceptance algorithm \citep{gale1962college, roth2008}. These mechanisms guarantee stability and incentive compatibility, but require centralized coordination and therefore are not directly applicable to decentralized markets such as the current job market in statistics and data science.

\section{Applications}
\label{sec:simulations}

\subsection{Design of Studies}
We evaluate the proposed market design using numerical studies calibrated to the statistics job market. We simulate $m=103$ departments and annual cohorts of $n=300$ candidates over 10 years with 200 independent replications, where $n$ approximates the number of new statistics PhDs entering the academic job market each year \citep{SED2024}, as described in Section~\ref{subsec:data_application}. Departments are constructed using peer assessment scores from \citet{usnews2026statistics} and institutional characteristics from the \citet{collegescorecard} College Scorecard. Based on these measures, departments are grouped into four tiers by rank: Tier~1 (ranks 1--10), Tier~2 (ranks 11--25), Tier~3 (ranks 26--50), and Tier~4 (ranks 51 and below).

Candidates are characterized by quality vectors $\mathbf{v}_i \in [0,1]^{p_\mathrm{v}}$ and preferences over department attributes elicited through the 15-item questionnaire described in Supplement~\sref{appendix:questionnaire}. Questionnaire responses are combined with department characteristics through a weighted similarity measure to produce alignment scores $f_j(\mathbf{q}_i, \mathbf{d}_j)\in[\tfrac{1}{2},1]$, which enter the construction of departmental utilities $U_{ji}$; details of this computation are provided in Supplement~\sref{appendix:utility_details}.
For participating candidates, alignment scores are directly observed; for non-participants, they are imputed using the worst-discloser rule~\eqref{eq:nondisclosure_rule}. Interview selection follows Algorithm~\ref{alg:confidence_calibrated_selection}, ranking candidates by estimated expected utility $\widehat{\mathcal{U}}_{ji}$ using $B=100$ bootstrap draws.
We vary the participation rate $\rho \in \{0\%, 5\%, 20\%, 50\%, 90\%, 100\%\}$, where $\rho$ denotes the proportion of candidates that submit the questionnaire.  To isolate the effect of preference information, we adopt a nested design in which the participant set at each $\rho$ is a subset of that at higher rates, ensuring that observed differences are driven solely by the additional preference signals from new participants.

\begin{figure}[t]
\centering
\includegraphics[width=\textwidth]{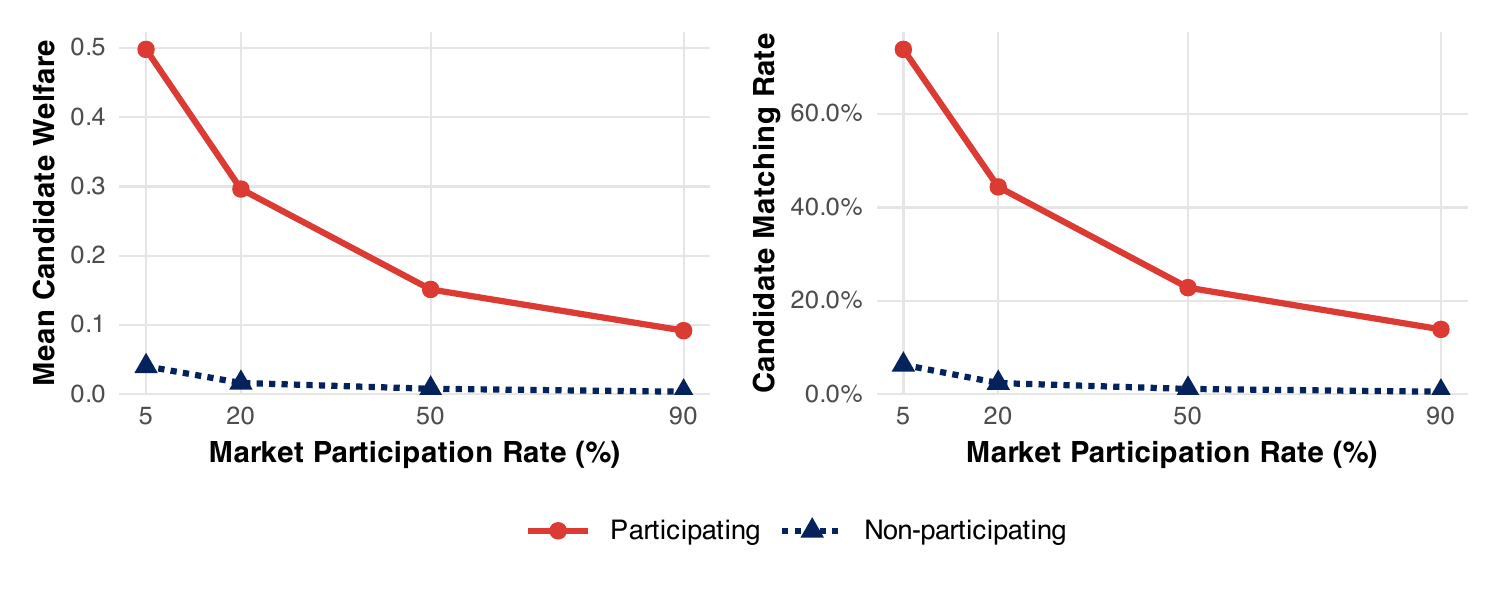}
\caption{Candidate welfare (left) and matching rate (right) by participation status across market participation rates $\rho$.}
\label{fig:candidate_by_participation}
\end{figure}

A 20-year burn-in period under the baseline $\rho=0\%$ generates historical data for training acceptance probability models in Section~\ref{sec:learning}.  In each year, a department is active with probability $0.6$, reflecting realistic hiring frequencies. Active departments interview $k_j=5$ candidates and extend one offer. In the first round, each department makes an offer to its top-ranked candidate, and candidates accept their most preferred offer. In a subsequent scramble round, departments with declined offers proceed down their interview lists.

\subsection{Candidate's Perspective}

\begin{figure}[t]
\centering
\includegraphics[width=0.85\textwidth]{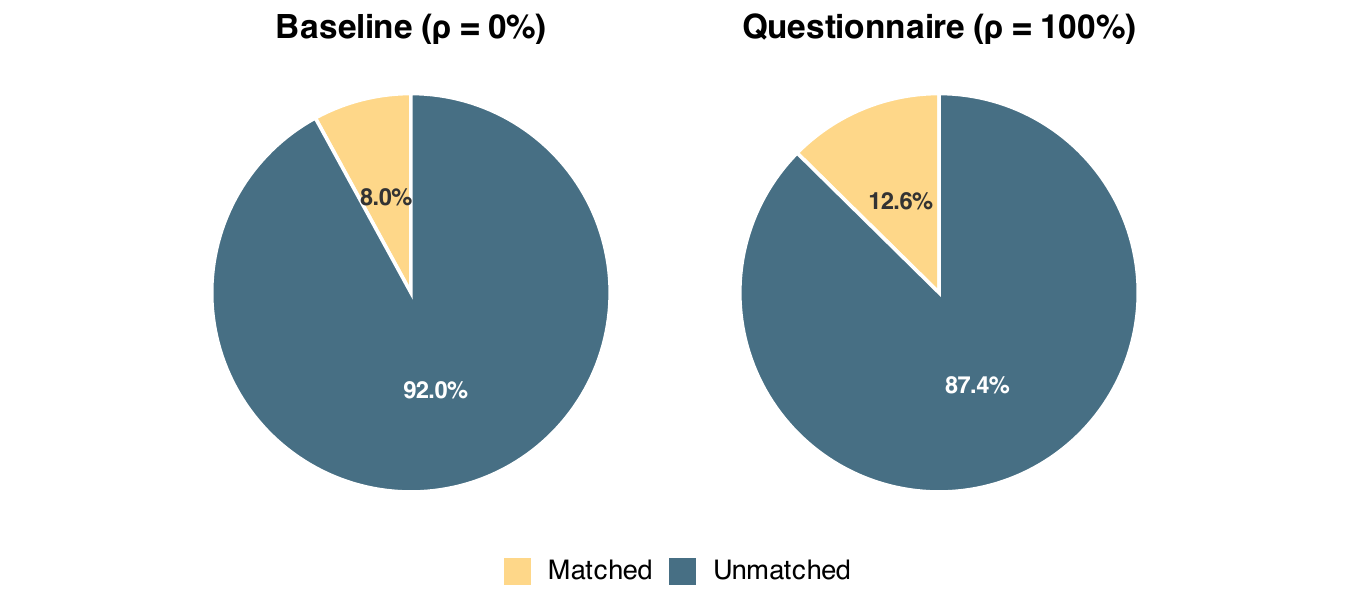}
\caption{Candidate outcome rates under baseline ($\rho=0\%$) and full participation ($\rho=100\%$).}
\label{fig:cand_outcome_pies}
\end{figure}

\paragraph{Incentives for Candidate’s Participation.}
Theorem~\ref{thm:disclosure} establishes that truthful participation strictly dominates non-participation. Figure~\ref{fig:candidate_by_participation} corroborates this result across all participation rates $\rho$: participating candidates consistently achieve higher welfare, measured by $W_i^{\mathcal{C}}$ in~\eqref{eqn:defwic}, as well as higher matching rates, defined as the proportion of candidates who secure a position. For example, at $\rho = 5\%$, early adopters attain more than 12 times the welfare of non-participants. As $\rho$ increases, the disadvantage of non-disclosure becomes more pronounced; at $\rho = 90\%$, non-participants have a matching rate of only $0.6\%$.
Figure~\ref{fig:cand_outcome_pies} further illustrates the improvement in candidate matching outcomes under the proposed questionnaire-based mechanism.

\begin{figure}[th]
\centering
\begin{subfigure}[t]{0.495\textwidth}
    \centering
    \includegraphics[width=\textwidth]{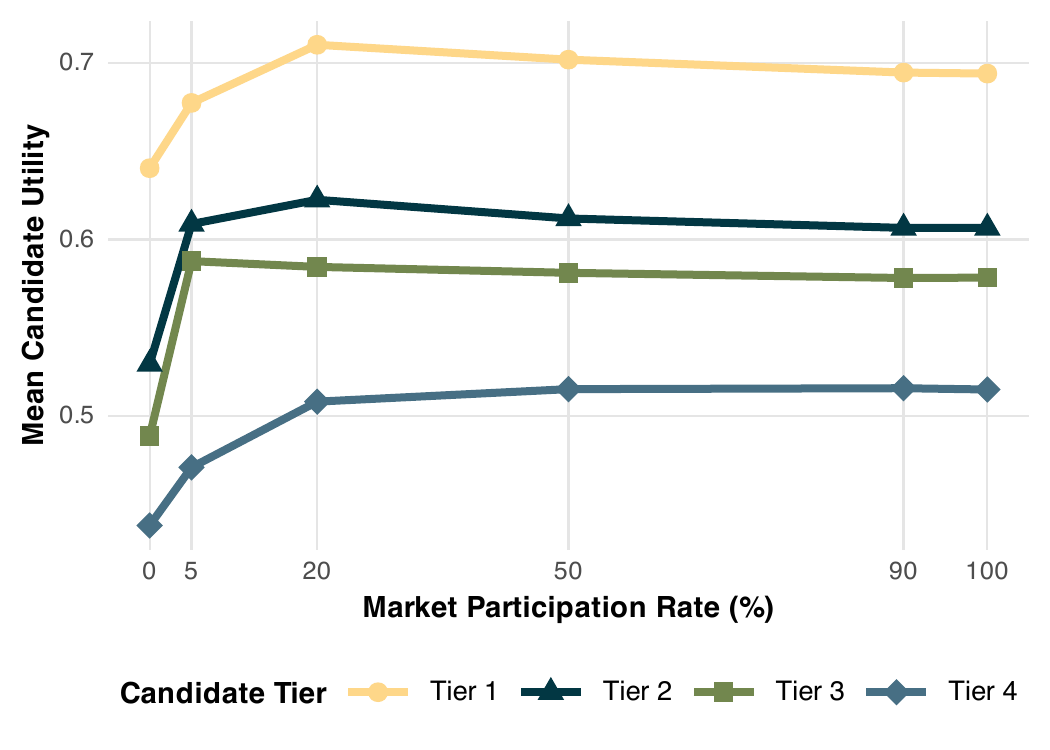}
    \caption{}
    \label{fig:candidate_utility_by_tier}
\end{subfigure}
\hfill
\begin{subfigure}[t]{0.495\textwidth}
    \centering
    \includegraphics[width=\textwidth]{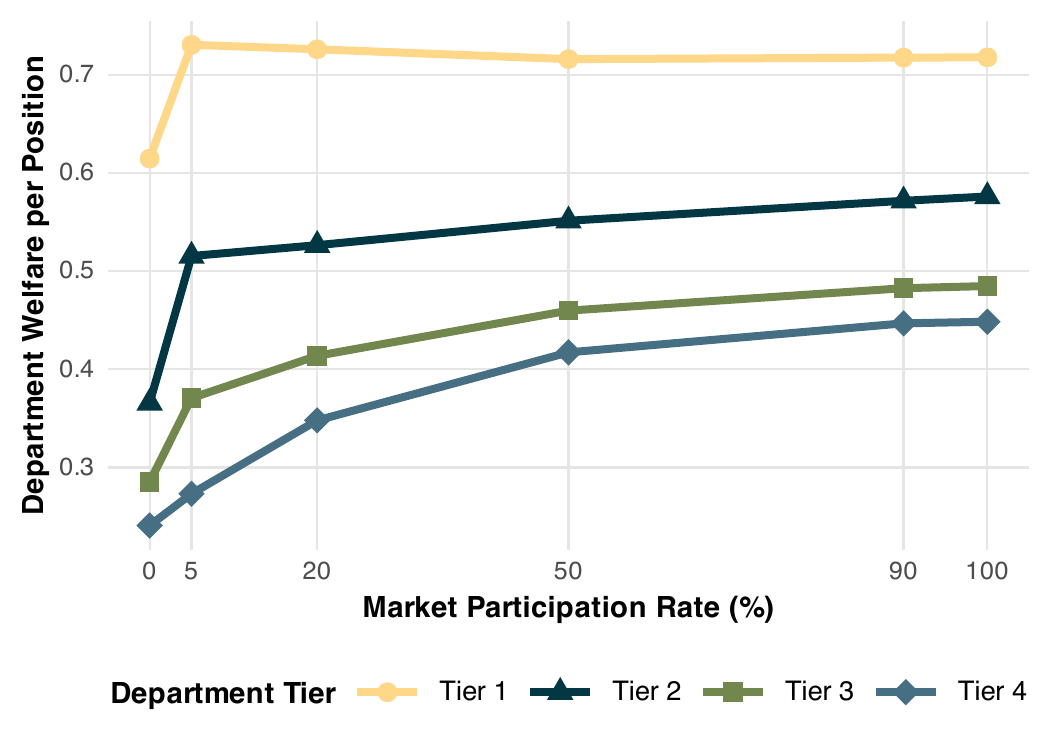}
    \caption{}
    \label{fig:department_welfare}
\end{subfigure}
\caption{Market outcomes by tier and participation rate $\rho$. (a) Mean candidate utility $\bar{V}_{ij}$ among matched candidates. (b) Department welfare per hiring position.}
\label{fig:welfare_combined}
\end{figure}

\paragraph{Heterogeneous Effects by Tier.}
Figure~\ref{fig:candidate_utility_by_tier} reports candidate utility $V_{ij} \in [0,1]$, defined in Section~\ref{subsec:candidate_incentives}, averaged over matched candidates within each quality tier.  Utility increases with participation across all tiers. Under $\rho = 0\%$, Tier~3 and Tier~4 candidates attain average utilities of $0.49$ and $0.44$, increasing to $0.58$ and $0.52$ under $\rho = 100\%$ (relative gains of $18\%$ and $18\%$), while Tier~1 and Tier~2 improve from $0.64$ and $0.53$ to $0.69$ and $0.61$ (relative gains of $8\%$ and $15\%$). These differences arise because, at $\rho = 0\%$, departments prioritize candidate quality and concentrate interviews among higher-tier applicants, so preference information yields the greatest marginal benefit for lower-tier candidates whose alignment would otherwise remain unobserved.

\subsection{Department's Perspective}

\paragraph{Department Welfare.}
As participation increases, departments observe alignment scores for a larger share of their applicant pool, enabling more informed interview selection. Figure~\ref{fig:department_welfare} shows that department welfare $W_j^{\mathcal{D}}$ in~\eqref{eqn:defwjd}, measured per hiring position, increases monotonically in $\rho$ across all tiers. Under $\rho = 0\%$, Tier~3 and Tier~4 departments attain welfare per position of $0.29$ and $0.24$, increasing to $0.48$ and $0.45$ under $\rho = 100\%$ (gains of $70\%$ and $86\%$), while Tier~1 departments improve from $0.61$ to $0.72$ ($17\%$). The larger gains for lower-tier departments reflect baseline inefficiencies, as these departments often interview high-quality candidates with low acceptance probabilities, resulting in rejected offers and unfilled positions. Preference signals shift interview selection toward candidates with stronger interest, reducing rejections and improving fill rates.

\begin{figure}[t]
\centering
\includegraphics[width=0.85\textwidth]{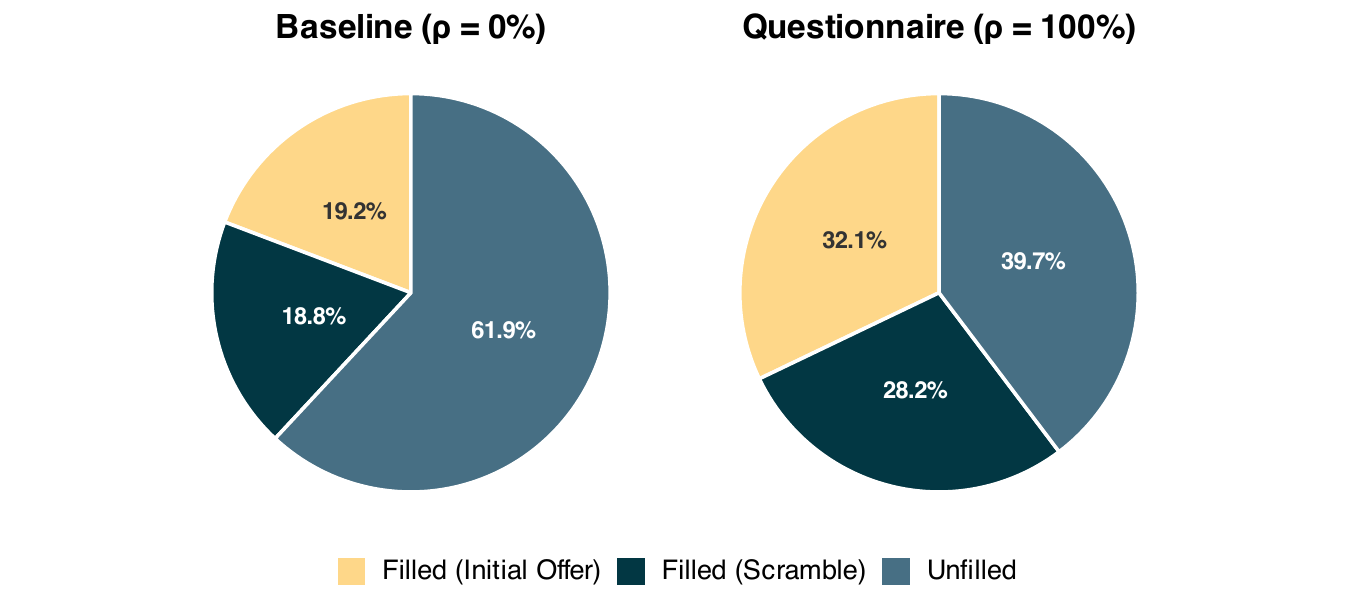}
\caption{Department position outcomes under baseline ($\rho=0\%$) and full participation ($\rho=100\%$).}
\label{fig:dept_outcome_pies}
\end{figure}

\begin{figure}[t]
\centering
\includegraphics[width=\textwidth]{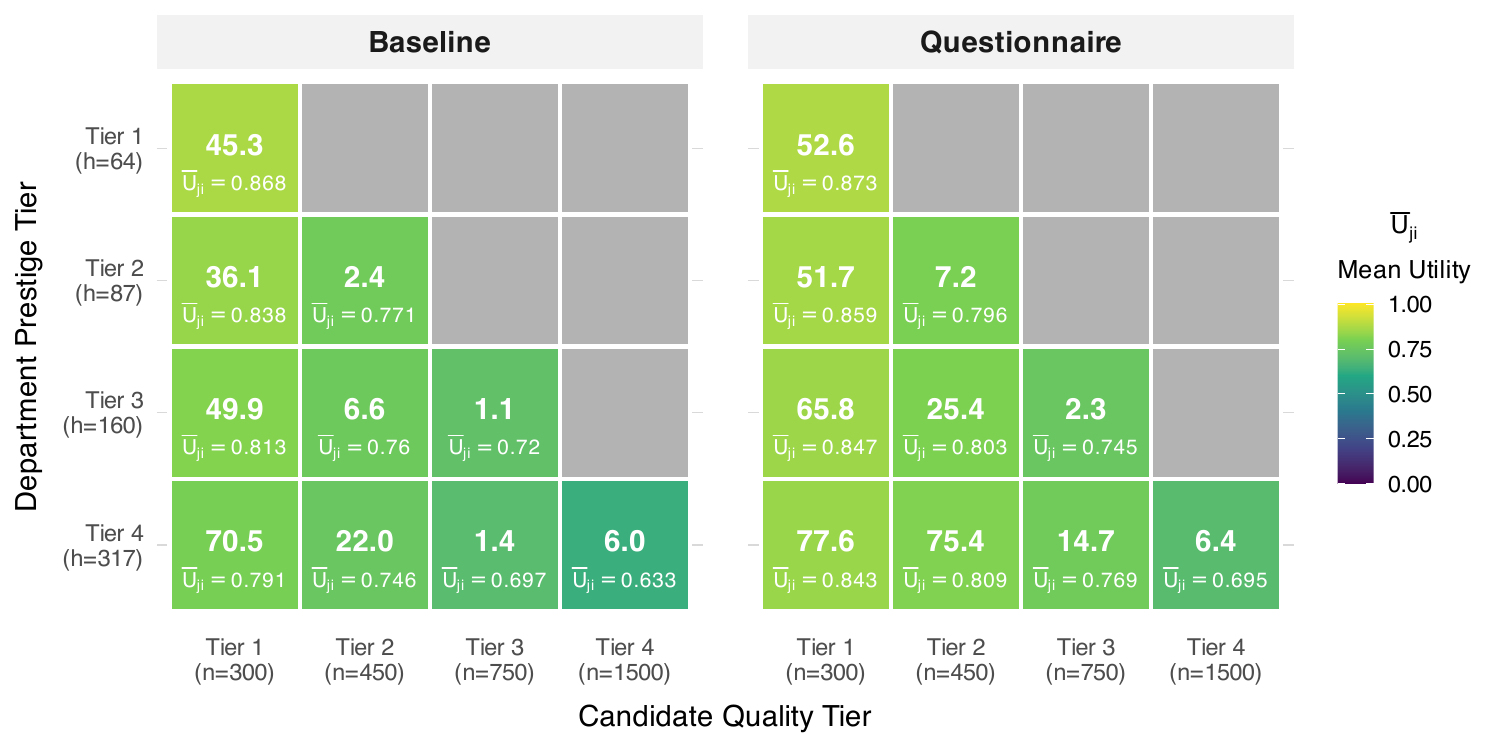}
\caption{Hiring distribution by department and candidate tier. Cell values show mean department utility $\bar{U}_{ji}$ and hire counts, averaged over 200 replications and aggregated across 10 years. Department tiers show total hiring quota $h$; candidate tiers show market size $n$. \textbf{Left:} Baseline ($\rho=0\%$). \textbf{Right:} Questionnaire ($\rho=100\%$).}
\label{fig:hiring_heatmap}
\end{figure}

\paragraph{Hiring Patterns.}
Figure~\ref{fig:dept_outcome_pies} shows the distribution of department position outcomes, distinguishing between positions filled in the first round, those filled in the scramble phase, in which departments with declined offers proceed down their interview lists, and those left unfilled. Under $\rho = 0\%$, $61.9\%$ of positions remain unfilled, with $19.2\%$ filled in the first round and $18.8\%$ in the scramble phase. Under $\rho = 100\%$, the unfilled rate decreases to $39.7\%$, while first-round hires increase to $32.1\%$ and scramble-phase hires to $28.2\%$. To understand the mechanisms underlying these improvements, Figure~\ref{fig:hiring_heatmap} shows the distribution of hires by department and candidate tier. Under $\rho=0\%$, lower-tier departments concentrate interviews on top-tier candidates with low estimated acceptance probabilities $\hat{\pi}_{ji}$ in~\eqref{eqn:estpi}, leading to frequent rejections and unfilled positions. For example, Tier~4 departments direct most interviews toward Tier~1 candidates, resulting in approximately $71$ hires on average from that tier with $\bar U_{ji} = 0.79$, where $\bar U_{ji}$ denotes the mean of the department utility $U_{ji}$ in~\eqref{utility} over the corresponding hires, and only $6$ hires from Tier~4 candidates. Under $\rho = 100\%$, these departments diversify their hiring to approximately $75$ Tier~2 and $15$ Tier~3 candidates, with $\bar U_{ji}$ of $0.81$ and $0.77$, respectively. With the questionnaire, departments reallocate interviews toward candidates with stronger alignment, producing a more balanced hiring pattern and increasing total hires.

\paragraph{Matching Stability.}
Theorem~\ref{thm:blocking_pairs} establishes that blocking pairs diminish as the alignment signal becomes more informative. As participation increases, the candidate-specific alignment scores $f_j(\mathbf{q}_i, \mathbf{d}_j)$ entering both $U_{ji}$ and $\hat{\pi}_{ji}$ reduce uncertainty in expected utility rankings, resulting in matches with stronger mutual fit and fewer blocking pairs. Supplement~\sref{appendix:utility_details} confirms this empirically, with the blocking pair rate, measured as the proportion of shortlist-eligible candidate--department pairs that form blocking pairs, decreasing monotonically from $81.8\%$ under $\rho = 0\%$ to $69.9\%$ under $\rho = 100\%$. 

\subsection{Pilot Design}
\label{subsec:pilot}

We propose a two-phase pilot study, illustrated in Figure~\ref{fig:overview_pilot}. Phase~1 develops a concise, standardized questionnaire to capture candidate-department fit, and Phase~2 evaluates the mechanism’s performance in interview selection and hiring outcomes under field conditions.

\paragraph{Phase~1: Questionnaire Development.}
The questionnaire is developed through an iterative consultation process with hiring departments. We begin by interviewing search committee members across a diverse set of institutions to identify preference dimensions that are most relevant for hiring decisions. These inputs are used to refine the questionnaire, guided by two types of design errors.
A Type~I error arises when the questionnaire includes items that do not improve the identification of well-matched candidates, thereby increasing respondent burden without adding informational value. A Type~II error arises when relevant dimensions of fit are omitted, reducing the effectiveness of the signal. Each iteration mitigates these errors by removing uninformative items and incorporating missing dimensions. 

\paragraph{Phase~2: Field Evaluation.}
The questionnaire is then implemented in a pilot study in which participating departments adopt the mechanism, while nonparticipating departments continue with standard hiring practices. 
Following the hiring cycle, outcomes at participating departments are compared with those at nonparticipating departments of similar standing. On the department side, we examine position fill rates and initial-offer acceptance rates. On the candidate side, we measure the proportion of candidates receiving interview invitations and the proportion ultimately hired. These outcomes provide an evaluation of the mechanism in improving interview targeting and hiring outcomes under field conditions.

\section{Related Works}
\label{sec:relatedworks}
We review related work from three areas: job market design, preference signaling, and matching markets.

\paragraph{Job Market Design.}
A large literature studies the design of job markets to reduce congestion and search costs. The redesign of the National Resident Matching Program showed how centralized mechanisms can stabilize large labor markets \citep{roth1999redesign}. Studies of the economics job market demonstrate substantial coordination failures, motivating proposals for improved coordination mechanisms \citep{coles2010job}. Recent designs across disciplines focus on improving early-stage screening and reducing logistical costs; for example, virtual and hybrid interviewing can expand participation while lowering financial costs \citep{termini2021using}. 
Our work differs from these approaches in two ways. First, we do not modify interview logistics or introduce a centralized matching mechanism. Second, we combine structured preference signaling with statistical ranking under uncertainty to improve the allocation of interview opportunities in job markets. 

\paragraph{Preference Signaling.}
Congestion in decentralized markets has motivated signaling mechanisms that allow candidates to communicate interest to a limited number of employers. In the economics job market, candidates can send a small number of preference signals, known as “roses,” to indicate strong interest in particular departments \citep{coles2013preference}. Such mechanisms can improve welfare by helping employers identify candidates with high interest \citep{ostrovsky2010information, ashlagi2020clearing}.
Our approach differs in two ways. First, instead of a small number of discrete signals, we introduce a standardized questionnaire that generates structured preference information. Second, the resulting signals are continuous and interpretable, allowing departments to infer candidate–department fit along multiple dimensions rather than relying on binary indicators of interest.

\paragraph{Matching Markets.}
Our work relates to the literature on two-sided matching markets, which studies how markets allocate agents with heterogeneous preferences. The foundational model of \citet{gale1962college} introduced the concept of stable matching, later applied to centralized allocation systems such as school choice and labor markets \citep{roth1984stability, abdulkadirouglu2003school}. A key insight from this literature is that decentralized markets often suffer from informational frictions and congestion, so that mutually beneficial matches may fail to occur \citep{dai2020multi,dai2020learning}.
Our framework connects to this literature by studying how improved information about candidate–department fit can improve outcomes in a decentralized job market. Unlike classical matching models that focus on the final matching stage or centralized mechanisms, we focus on the earlier interview-selection stage. We introduce a preference-signaling mechanism together with a learning-based ranking procedure that uses historical data to guide interview allocation.

\section{Conclusion}
\label{sec:conclusion}
This paper proposes a market design framework to improve efficiency and coordination in the academic job market for new faculty hires in statistics and data science. The framework introduces a centralized signaling mechanism based on a standardized questionnaire that allows candidates to communicate structured preference information while preserving decentralized hiring decisions. Departments then combine these signals with traditional application materials and data-driven estimates of acceptance probabilities to guide interview selection.
The theoretical analysis establishes key properties of the mechanism, including incentives for candidate participation and stability of the resulting allocation when preference information becomes sufficiently informative. Simulation studies calibrated to realistic market conditions demonstrate substantial welfare gains and improved candidate–department alignment relative to current decentralized practices without signaling.
The results show that structured preference signaling combined with statistically principled ranking can mitigate congestion and coordination failures in academic job markets. The proposed framework is simple to implement and can be integrated into existing professional platforms.

Two directions for future research are particularly promising. First, empirical implementation and pilot experiments could evaluate the performance of the proposed mechanism in real hiring cycles. Second, the framework could be extended to incorporate richer forms of learning, such as strategic interactions among departments, allowing the mechanism to adapt to evolving market conditions over time.

\bibliographystyle{chicago}
\bibliography{match}
\end{document}